\documentclass[twocolumn]{aastex631}

\usepackage{amsmath}
\usepackage{amssymb}
\usepackage{graphicx}
\usepackage{bm}

\newcommand{\fNL}{f_\mathrm{NL}}
\newcommand{\fNLloc}{f_\mathrm{NL}^\mathrm{local}}
\newcommand{\fNLeq}{f_\mathrm{NL}^\mathrm{equil}}
\newcommand{\fNLort}{f_\mathrm{NL}^\mathrm{ortho}}

\newcommand{\kmax}{k_\mathrm{max}}

\def\bk{\mathbf{k}}

\def\bx{\mathbf{x}}
\def\d{\mathrm{d}}

\newcommand{\hMpc}{h\,\mathrm{Mpc}^{-1}}

\def\btheta{\boldsymbol{\theta}}
\def\t{\mathbf{t}}
\def\cov{\mathbf{C}}
\def\mean{\boldsymbol{\mu}}
\def\data{\mathbf{d}}
\def\transpose{\mathrm{T}}
\def\fisher{\mathbf{F}}
\newcommand{\Quijote}{\textsc{Quijote}}
\newcommand{\QuijotePNG}{\textsc{Quijote-png}}

\newcommand{\cnrs}{Sorbonne Universit\'{e}, CNRS, UMR 7095, Institut d'Astrophysique de Paris, 98 bis bd Arago, 75014 Paris, France}
\newcommand{\cca}{Center for Computational Astrophysics, Flatiron Institute, 162 5th Avenue, New York, NY 10010, USA}
\newcommand{\bologna}{Dipartimento di Fisica e Astronomia, Alma Mater Studiorum - University of Bologna, Via Piero Gobetti 93/2, 40129 Bologna BO, Italy}
\newcommand{\inaf}{INAF - Osservatorio Astronomico di Bologna, Via Piero Gobetti 93/3, 40129 Bologna BO, Italy}
\newcommand{\infn}{INFN - Istituto Nazionale di Fisica Nucleare, Sezione di Bologna, Viale Berti Pichat 6/2, 40127 Bologna BO, Italy
}
\newcommand{\infnPad}{INFN, Sezione di Padova, via Marzolo 8, I-35131, Padova, Italy}
\newcommand{\mpa}{Max-Planck-Institut f\"ur Astrophysik, Karl-Schwarzschild-Straße 1, 85748 Garching, Germany}
\newcommand{\ICC}{ICC, University of Barcelona, IEEC-UB, Martí i Franquès, 1, E-08028 Barcelona, Spain}
\newcommand{\ICREA}{ICREA, Pg. Lluís Companys 23, Barcelona, E-08010, Spain}
\newcommand{\Galilei}{Dipartimento di Fisica e Astronomia “G. Galilei”,Università degli Studi di Padova, via Marzolo 8, I-35131, Padova, Italy}
\newcommand{\uwc}{Department of Physics and Astronomy, University of the Western Cape, Cape Town 7535, South Africa}
\newcommand{\princeton}{Department of Astrophysical Sciences, Princeton University, 4 Ivy Lane, Princeton, NJ 08544 USA}

\begin{document}

\title{Quijote-PNG: Quasi-maximum likelihood estimation of Primordial Non-Gaussianity in the non-linear halo density field}

\author{Gabriel Jung}
\affiliation{\Galilei}
\affiliation{\infnPad}
\author{Dionysios Karagiannis}
\affiliation{\uwc}
\author{Michele Liguori}
\affiliation{\Galilei}
\affiliation{\infnPad}
\author{Marco Baldi}
\affiliation{\bologna}
\affiliation{\inaf}
\affiliation{\infn}
\author{William R Coulton}
\affiliation{\cca}
\author{Drew Jamieson}
\affiliation{\mpa}
\author{Licia Verde}
\affiliation{\ICREA} 
\affiliation{\ICC}
\author{Francisco Villaescusa-Navarro}
\affiliation{\cca}
\affiliation{\princeton}
\author{Benjamin D. Wandelt}
\affiliation{\cnrs}
\affiliation{\cca}

\begin{abstract}
We study primordial non-Gaussian signatures in the redshift-space halo field on non-linear scales, using a quasi-maximum likelihood estimator based on optimally compressed power spectrum and modal bispectrum statistics. We train and validate the estimator on a suite of halo catalogues constructed from the \QuijotePNG\ N-body simulations, which we release to accompany this paper. We verify its unbiasedness and near optimality, for the three main types of primordial non-Gaussianity (PNG): local, equilateral, and orthogonal. We compare the modal bispectrum expansion with a $k$-binning approach, showing that the former allows for faster convergence of numerical derivatives in the computation of the score function, thus leading to better final constraints. We find, in agreement with previous studies, that the local PNG signal in the halo field is dominated by the scale-dependent bias signature on large scales and saturates at $k \sim 0.2~\hMpc$, whereas the small-scale bispectrum is the main source of information for equilateral and orthogonal PNG. Combining power spectrum and bispectrum on non-linear scales plays an important role in breaking degeneracies between cosmological and PNG parameters; such degeneracies remain however strong for equilateral PNG. We forecast that PNG parameters can be constrained with $\Delta \fNLloc = 45$, $\Delta \fNLeq = 570$, $\Delta \fNLort = 110$, on a cubic volume of $1 \left({ {\rm Gpc}/{ {\rm h}}} \right)^3$, at $z = 1$, considering scales up to $\kmax = 0.5~\hMpc$.

\end{abstract}


\section{Introduction}
\label{sec:introduction}

The coming generation of spectroscopic and photometric galaxy surveys -- e.g., Euclid, DESI, Spherex, Rubin Observatory, Roman \citep{Euclid_2011, DESI_2016, Dore_2013,LSST_2009} -- will allow us to study galaxy clustering with an unprecedented level of accuracy and precision, shedding further light on many open questions in cosmology.
Among the many exciting possibilities, an interesting prospect, which we mainly focus on in this work, will be that of improving our understanding of Early Universe physics, via high precision tests of Primordial non-Gaussianity (PNG). 

Cosmic Microwave Background (CMB) measurements \citep{Planck:2019kim}, in agreement with theoretical expectations, have constrained the primordial cosmological perturbation field to be at most weakly non-Gaussian. This implies, for a large majority of Early Universe scenarios, that most of the PNG information is contained in the primordial bispectrum. 
For this reason, the bispectrum of dark matter tracers (e.g., galaxies) in Large Scale Structure (LSS) can be a powerful probe of PNG. Crucially, the 3D galaxy bispectrum gives us also access, in principle, to a larger number of modes with respect to the 2D (angular) bispectrum of CMB anisotropies. Therefore, LSS bispectrum analyses can potentially lead to significant improvements in PNG constraints over current, CMB-based, results. Achieving such improvements will require however to include non-linear scales in the analysis, carrying strong non-Gaussian (NG) signatures which are not primordial, but arise from late-time, non-linear evolution of cosmic structures. Disentangling the NG late time component from the subdominant primordial one is therefore a crucial challenge in this kind of studies. It can be addressed either by analytical modeling of the bispectrum -- via a suitable perturbative approach at mildly non-linear scales \citep{Cabass:2022wjy, Cabass:2022ymb, DAmico:2022gki} -- or by relying on fully numerical approaches, which evaluate the bispectrum  \citep[and/or other summary statistics, see, e.g.,][]{Biagetti:2020skr,Friedrich_2020, wavelets} using large mock datasets; field-level inference on large scales, not relying on specific statistical summaries, has also been recently considered, see \citet{Andrews:2022nvv}. In this work, which is the fourth in a series of papers, following \citet{Jung:2022rtn,Coulton:2022qbc,Coulton:2022rir}, we base our study on the \QuijotePNG\ simulation suite, recently presented in \citet{Coulton:2022qbc}. Our main goal is to quantify the accuracy with which $\fNL$ can be constrained using both the power spectrum and bispectrum of the dark matter halo field, up to strongly non-linear scales ($\kmax = 0.5~\hMpc$), using simulations with different kinds of PNG. More precisely, we consider the three main PNG bispectrum shapes, namely, the local, equilateral, and orthogonal shapes, which are predicted in a large variety of inflationary scenarios. 

This work extends our initial analysis presented in \citet{Jung:2022rtn}, where we worked at the level of the matter field. As in the previous analysis, we derive forecasts for PNG and standard cosmological parameters, by combining power spectrum and bispectrum measurements at non-linear scales; our main focus is then on building an unbiased and nearly optimal quasi-maximum likelihood estimator, based on applying a MOPED-like compression algorithm to a modal decomposition of the data bispectrum. 
In our companion paper \citep{Coulton:2022rir} we independently perform a similar analysis at a different redshift \citep[$z=0$ in][vs.\ $z=1$ in this work]{Coulton:2022rir}, but we employed a binned decomposition in $k$-space instead of the modal approach adopted here; in that work we studied in detail the PNG information content in the halo field while focusing on important numerical convergence issues. Therefore, the two analyses are complementary to study the robustness of our approach and to cover a full range of crucial issues, from numerical stability, to the precise quantification of the information gain obtained from different observables at different scales and the demonstration of nearly optimal and unbiased bispectrum data compression for parameter estimation. Taken together, we think these works represent an important development in the effort to build a data analysis pipeline to be applied to observations. We release the halo catalogues of the \QuijotePNG\footnote{\url{https://quijote-simulations.readthedocs.io/en/latest/png.html}} suite used in these works.
 
Since in the current work we consider tracers of the underlying density field, an additional signature of PNG arises -- in comparison to the previous matter field analysis -- in the form of a scale-dependency in the tracer bias. Such feature has a power law behaviour, with degree determined by the squeezed limit of the PNG bispectrum shape under study: it is most prominent for local NG, with a $\sim {1/k^2}$ behaviour and absent in the equilateral case. Scale-dependent bias has been the object of significant study in the literature, see, e.g., \citet{Dalal2008, Matarrese2008, Slosar2008, Afshordi2008, Seljak_2009, Desjacques2010, Castorina_2018, Chan_2019, Giri_2022}, and was used to extract local PNG constraints from BOSS data \citet{Slosar2008,Ross_2013,Leistedt_2014,Mueller_2021,Cabass:2022wjy,DAmico:2022gki}. Recently, it has been however pointed out that accurate modeling of scale-dependent bias from PNG also depends on details of galaxy formation, making its use as a tool to measure the PNG amplitude $\fNL$ significantly more challenging than previously thought \cite{Barreira:2020ekm,Barreira:2021ueb,Barreira:2022sey}.
The effect of scale-dependent bias is automatically incorporated in our analysis, where we are mostly concerned with assessing its relative constraining power on different NG shapes, as compared to the bispectrum, and verifying agreement with both our analysis in \citet{Coulton:2022rir} and theoretical expectations \citep[see, e.g.][]{dePutter_2018,Karagiannis:2018jdt}.

The paper is structured as follows. In section \ref{sec:shapes} we briefly review the NG models considered in the analysis. In section \ref{sec:method} we illustrate our methodology for data compression and parameter estimation. In section \ref{sec:analyses} we discuss our Fisher matrix analysis, showing expected parameter constraints on different scales, and describe the application of our quasi-maximum likelihood, joint power spectrum and bispectrum estimator to simulated data. In section \ref{sec:conclusion} we summarize our main results and draw our final conclusions. Finally, in appendix \ref{app:shot-noise} we provide more details about the implementation of shot-noise modes in the bispectrum estimator. In appendix \ref{app:binned} we show a comparison between the modal and binned approaches to bispectrum estimation, and in appendix \ref{app:carpool} we discuss the results of a preliminary study aimed at the application of the CARPool technique to the evaluation of covariances and numerical derivatives.

\section{Bispectrum shapes}
\label{sec:shapes}

Violating any condition of the standard inflationary model induces a deviation from the perfect Gaussian initial conditions, which leads to non-zero high-order correlators. The largest of them, in most inflationary models, is the bispectrum, i.e. the three-point correlation function of Fourier modes, defined as:
\begin{equation}
\begin{split}
\langle \Phi(\mathbf{k_1}) \Phi(\mathbf{k_2}) \Phi(\mathbf{k_3}) \rangle = (2 \pi)^3 & B_\Phi(k_1, k_2, k_3) \\ &\times \delta^{(3)}(\mathbf{k_1} + \mathbf{k_2} + \mathbf{k_3}).
\end{split}
\end{equation}
The primordial bispectrum is generally written as 
\begin{equation}\label{eq:Bphi}
B_\Phi(k_1, k_2, k_3) = \fNL F(k_1, k_2, k_3),
\end{equation}
where $\fNL$ is the dimensionless amplitude parameter corresponding to a given primordial bispectrum shape $F(k_1, k_2, k_3)$, which encompasses the dependence of the bispectrum on triplets of Fourier space modes.

In this work, we focus on building estimators to measure $\fNL$ of three of the most common primordial shapes, namely the local, equilateral and orthogonal\footnote{We use the orthogonal-LSS shape described in \citet{Coulton:2022qbc}, which is a better approximation of the non-separable orthogonal bispectrum than the standard template of CMB analyses for the 3D matter field.} bispectra \citep[see][and references therein for the complete description of these templates]{Coulton:2022qbc}.

For dark matter tracers (e.g. halos), the presence of PNG has a significant impact, due to the introduced coupling between large and small scale modes, with the most known example being that of the local type. In this case, the halo overdensity on large scales will no longer depend only on the matter overdensity, but also on the primordial gravitational potential \citep[see][for a review]{Desjacques2016}. This results in a scale-dependent term that introduces an important PNG signature on the large scales of a correlator. This is of particular importance to the power spectrum of the observed dark matter tracers, since it enhances the PNG signal within the two-point correlation function, which otherwise would have been very limited, e.g. in the case of a dark matter field analysis \citep[see e.g.][]{Coulton:2022qbc,Jung:2022rtn}.

The effect of the scale dependent term on the power spectrum has been extensively studied in the literature, especially for the local PNG type. However, recent developments have made the measurement of $\fNL$, by such a term in the power spectrum, challenging, due to the perfect degeneracy between $\fNL$ and the scale dependent bias coefficient $b_\phi$ \citep{Barreira:2020ekm,Barreira:2022sey}. For the halo bispectrum, a significant amount of the PNG signal is located within the primordial part (eq.~\ref{eq:Bphi}), while the scale dependent terms, studied at a theoretical level e.g. in \citet{Karagiannis:2018jdt}, that could carry a notable amount of information on local PNG, suffer from the same limitations as the power spectrum \citep{Barreira:2021ueb}.

The effect of the scale-dependent bias will be taken into account within the framework of the forward-modeling. In a simulation-based approach we assume tight priors on the scale dependent bias coefficient $b_\phi$, in order to focus on the $\fNL$ constraints \citep[see also][for an extensive discussion]{Coulton:2022rir}.

\section{Method}
\label{sec:method}

In this section we review the main aspects of our methodology for data compression and quasi-maximum likelihood estimation of cosmological and PNG parameters, starting from the evaluation of power spectrum and modal bispectrum summary statistics.

\subsection{Quasi maximum-likelihood estimator}
\label{sec:compression}

Starting from a given data vector $\data$ (a given set of summary statistics, like the power spectrum and/or the bispectrum) that depends on some parameters of interest denoted $\btheta$ (e.g.\ $\fNL$), one can write the following quasi maximum-likelihood estimator for the value of the parameters \citep[see][for details]{Alsing:2017var}:
\begin{equation}
    \hat\btheta = \btheta_* + \fisher^{-1}_*\left[\nabla_{\btheta}\mean_*^\transpose\cov^{-1}_*(\data - \mean_*)\right] \equiv \btheta_* + \fisher^{-1}_*\t,
    \label{eq:estimator}
\end{equation}
where the subscript $*$ denotes that the quantities are evaluated at some chosen fiducial point, and $\mean$ and $\cov$ are, respectively, the mean and the covariance of $\data$. The two key ingredients of this estimator, which are the Fisher information $\fisher$ and the compressed score statistic $\t$, will be detailed below. Note also that in this expression we assume a Gaussian likelihood and a dependence on parameters through the mean only, a reasonable assumption as verified in \citet{Jung:2022rtn}.

The Fisher matrix, a standard method to evaluate the information content of some observables, is given by:
\begin{equation}
    \fisher = \nabla_{\btheta}\mean^\transpose \cov^{-1} \nabla^\transpose_{\btheta}\mean.
    \label{eq:fisher}
\end{equation}
This requires knowledge of the derivatives $\nabla_{\btheta}\mean$ and the covariance $\cov$, which can be both evaluated from a large set of simulations, as we do in this work (see section~\ref{sec:analyses}). However, reaching numerical convergence for the joint analysis of multiple parameters may be very challenging and therefore prone to wrong results. In \citet{Jung:2022rtn} we checked, by studying the matter field, that if the covariance matrix is not converged, it would typically induce suboptimal error bars, and that  non-converged derivatives could bias the estimated parameters. In \citet{Coulton:2022rir} we showed that noisy derivatives could lead to overconfident error bars when working with the halo field.

To tackle this problem, the alternative compressed Fisher method described in \citet{Coulton:2022rir} \citep[see also][for details]{Coulton_2022b} can provide conservative bounds. It consists of two steps, the first of which is to compress the data to the score function using \citep[see][]{Alsing:2017var}
\begin{equation}
    \t = \nabla_{\btheta}\mean_*^\transpose\cov^{-1}_*(\data - \mean_*)~,
    \label{eq:compression}    
\end{equation}
that is equivalent to the MOPED compression scheme of \citet{Heavens:1999am}. This operation reduces the data vector $\data$ of size $n$ down to only $p$ numbers, where $p$ is the number of parameters of interest, while keeping all relevant information about these parameters. Then, to compute the compressed Fisher matrix one has only to apply the standard expression of eq.\ \eqref{eq:fisher} to the compressed data. An important subtlety of this scheme is that it requires to use two separate sets of simulations for the two different steps. The first part is used for the compression step, to build a new summary statistics, which will be suboptimal if derivatives are noisy. The second part is then compressed and used to estimate the Fisher matrix from the compressed statistics, which is suboptimal if the compression step is suboptimal, but is also a lot less noisy due to the much lower dimensionality of the compressed statistics.\footnote{The variance of the procedure can be significantly decreased by repeating the procedure many times (generating different splittings each time) and computing the Monte Carlo average of the compressed Fisher matrix.}

\subsection{Summary statistics}
\label{sec:statistics}

In this work, we use the same observables as in \citet{Jung:2022rtn}, based on the power spectrum and bispectrum statistics as they contain significant and complementary information about both the $\Lambda$CDM cosmological parameters and the PNG amplitudes $\fNL$.

The standard power spectrum estimator of a field $\delta(\bk)$ defined on a grid of fundamental mode $k_f$ is given by
\begin{equation}
    \label{eq:P(k)}
    \hat{P}(k_i) = \frac{1}{ V N_i} \sum\limits_{\bk \in \Delta_i} \delta(\bk)\delta^*(\bk),
\end{equation}
where $V$ is the survey volume, and a binning of the $k$-range has been introduced with each bin $\Delta_i$ having a width $k_f$ and containing $N_i$ independent vectors of $\bk$.

As was initially shown for CMB NG analysis in \citet{Fergusson:2009nv, Fergusson:2010dm}, and extended later to LSS in \citet{Fergusson:2010ia, Regan:2011zq, Schmittfull:2012hq} \citep[see also][]{Lazanu:2015rta, Lazanu:2015bqo,Hung:2019ygc, Hung:2019nma, Byun:2020rgl, Byun:2022rvn}, the bispectrum information can be efficiently extracted from data by measuring the following modal coefficients:
\begin{equation}
    \label{eq:beta}
    \hat{\beta}_n = \frac{1}{V}\int \d^3 x\, M_p(\bx)M_q(\bx)M_r(\bx),
\end{equation}
where
\begin{equation}
    \label{eq:M}
    M_p(\bx) \equiv \int \frac{\d^3 k}{(2\pi)^3} \frac{q_p(k)\delta(\bk)}{\sqrt{k P(k)}}e^{i\bk.\bx},
\end{equation}
for a well-chosen basis of one-dimensional functions $q_p(k)$ and mode triplets $n \leftrightarrow (p,q,r)$. We refer the reader to \citet{Jung:2022rtn} for the details of the exact setup we use for the analyses presented in section~\ref{sec:analyses}, as they are almost identical (the only change is the addition of two special modes, introduced in \citet{Byun:2020rgl} and recalled in appendix~\ref{app:shot-noise}, describing the shot-noise component of the bispectrum expected from halos).

\section{Analyses}
\label{sec:analyses}

\subsection{Specifications}
\label{sec:specifications}

\begin{deluxetable*}{c|cccccccc}
\tablecaption{The fiducial values of the cosmological parameters and PNG amplitudes, together with their variations, used in the analysis.  \label{tab:quijote}}
\tablehead{& $\sigma_8$ & $\Omega_m$ & $n_s$ & $h$ & $M_\mathrm{min} (M_\odot/h)$ & $\fNLloc$ & $\fNLeq$ & $\fNLort$}
\startdata
Fiducial & 0.834 & 0.3175  & 0.9624 & 0.6711 & $3.2 \times 10^{13}$ & 0 & 0 & 0 \\
    Steps & $\pm 0.015$ & $\pm 0.01$ & $\pm 0.02$ & $\pm 0.02$ & $\pm 0.1 \times 10^{13}$ &  $\pm 100$ & $\pm 100$ & $\pm 100$ \\
\enddata
\end{deluxetable*}

For our analysis we use the publicly available \Quijote\footnote{\url{https://quijote-simulations.readthedocs.io}} and \QuijotePNG\footnote{\url{https://quijote-simulations.readthedocs.io/en/latest/png.html}} suites of N-body simulations \citep{Villaescusa-Navarro:2019bje, Coulton:2022qbc}. Each simulation represents a periodic cubic box of length $1~h^{-1}$Gpc, which contains $512^3$ particles, run with the \textsc{Gadget-III} code \citep{Springel:2005mi}. Initial conditions are generated at $z_i=127$ with the codes \textsc{2LPTIC} \citep{Crocce:2006ve} in the Gaussian case, and \textsc{2LPTPNG}\footnote{\url{https://github.com/dsjamieson/2LPTPNG}} in the non-Gaussian case \citep{Scoccimarro:2011pz, Coulton:2022qbc}; linear matter power spectra and transfer functions are obtained from CAMB \citep{Lewis:1999bs}. Finally, dark matter halos are identified using the Friends-of-Friends algorithm \citep{1985ApJ...292..371D} with a value of the linking length equal to $b=0.2$; we select those with a mass larger than $M_\mathrm{min}=3.2 \times 10^{13} M_\odot/h$, corresponding to a number density $\bar{n}\sim 5.1 \times 10^{-5} h^3\,\mathrm{Mpc}^{-3}$ at $z=1$ \citep[see][for a power spectrum and bispectrum analysis of these halo catalogues focused on cosmological parameters]{Hahn:2019zob}. 

We construct the halo density field in redshift-space at $z=1$ by depositing the halo positions, displaced radially by the velocity, on a grid of size $N_\mathrm{grid}=256$, using a fourth-order interpolation scheme implemented in the \textsc{Pylians3} code\footnote{\url{https://github.com/franciscovillaescusa/Pylians3}} \citep{Pylians}. We then measure the power spectrum and modal bispectrum monopoles using the estimators \eqref{eq:P(k)} and \eqref{eq:beta} including modes up to $\kmax=0.5~\hMpc$.

For the numerical computation of the covariance matrix, we have $15,000$ simulations at fiducial cosmology, whereas smaller sets of $500$ realizations, with varying input parameters, are used to evaluate the derivatives in eq.\ \ref{eq:compression}. In particular the analysis is focused on the PNG amplitudes of the three shapes considered here, $f_\mathrm{NL}^\mathrm{local}$, $f_\mathrm{NL}^\mathrm{equil}$, $f_\mathrm{NL}^\mathrm{orth}$; four cosmological parameters $\Omega_m$, $n_s$, $\sigma_8$ and $h$; and one parameter related to the halo bias, $M_\mathrm{min}$. The variation of the minimum halo mass generates distinct catalogs with $M_\mathrm{min}^{\rm fid}\pm \Delta M_\mathrm{min} $ (see table~\ref{tab:quijote}), which consequently leads to a variation of the halo number density. This is roughly equivalent to a variation of the linear bias contribution \citep[see e.g.][for details]{Desjacques2016}, while it propagates, to a minor extent, to higher order terms. Although, this bias model is quite simplistic, it is still useful within the framework of a first-order analysis presented in this work. A thorough investigation on the impact of the bias parameters, within a simulation-based approach, requires the population of halos with a HOD and the variation of the HOD parameters, which is left for future work. More details about the specifications of these simulations, concerning all the parameters considered in our analyses, can be found in table~\ref{tab:quijote}.

\subsection{Fisher constraints}
\label{sec:fisher}

\begin{figure*}
    \includegraphics[width=0.99\linewidth]{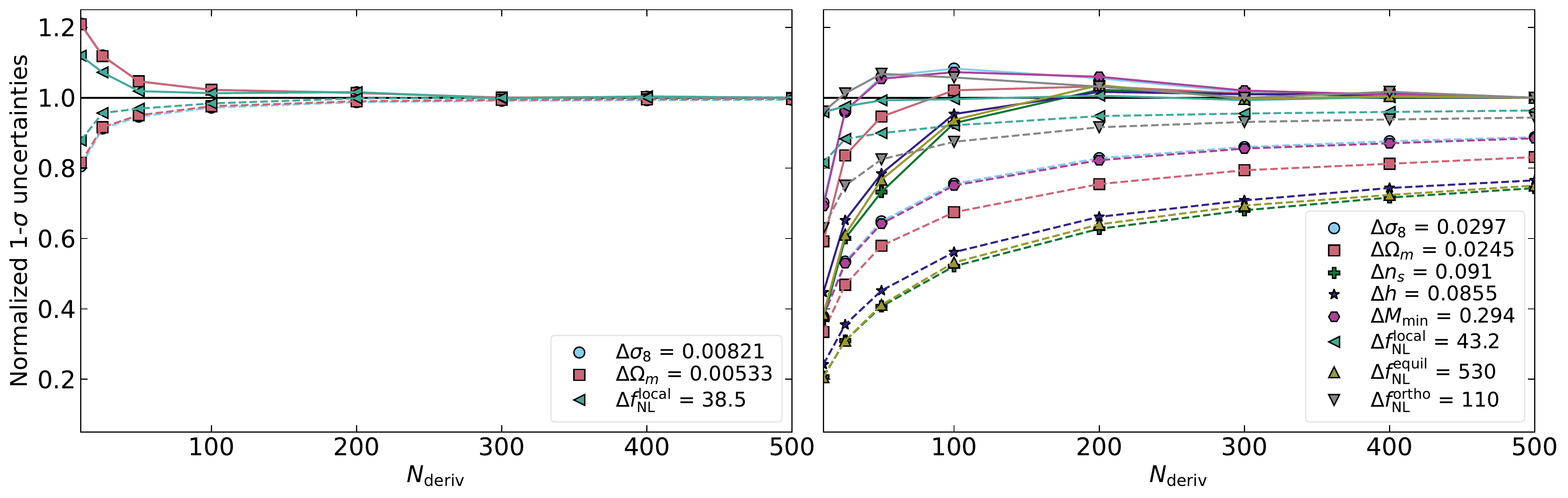}
    \caption{The stability of the Fisher $1$-$\sigma$ uncertainties under variations of the number of simulations used to compute derivatives. The analysis includes both the power spectrum and the bispectrum information of the halo field at $z=1$, with scales up to $\kmax=0.5~\hMpc$.
    In the left panel, three parameters $\{\sigma_8, \Omega_m, \fNLloc \}$ are analyzed jointly, while in the right panel $\{h, n_s, M_\mathrm{min}, \fNLeq, \fNLort\}$ are also included. The dashed lines correspond to the standard Fisher error bars and the solid lines are computed from the compressed summary statistics (see section~\ref{sec:compression}).}
    \label{fig:fisher_compressed}
\end{figure*}

\begin{figure*}
    \includegraphics[width=0.99\linewidth]{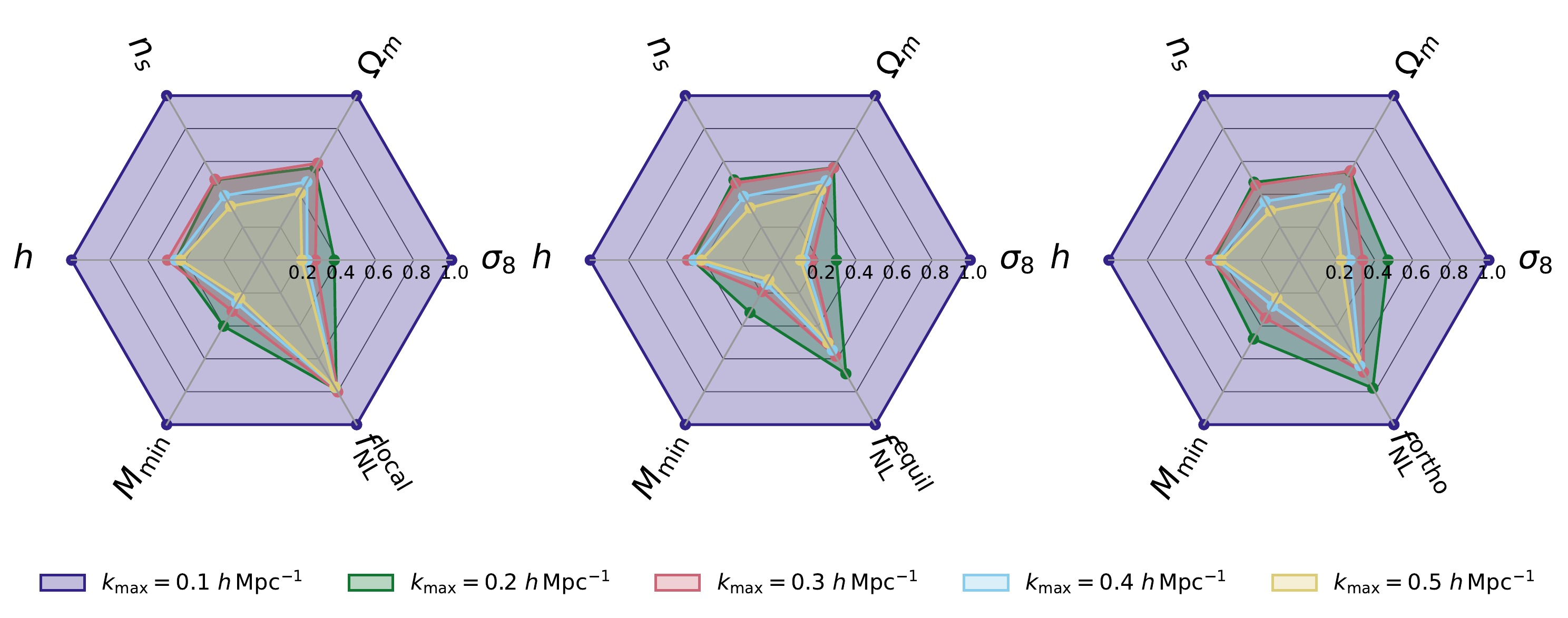}
    \caption{Normalized $1$-$\sigma$ Fisher error bars for the joint analysis of cosmological parameters and one PNG shape at a time, using both the power spectrum and modal bispectrum information of the halo field at $z=1$ for different $\kmax$ from $0.1~\hMpc$ to $0.5~\hMpc$. All error bars are computed from the compressed summary statistics (see section~\ref{sec:compression}).}
    \label{fig:fisher_pk_bisp}
\end{figure*}

\begin{figure*}
    \includegraphics[width=0.99\linewidth]{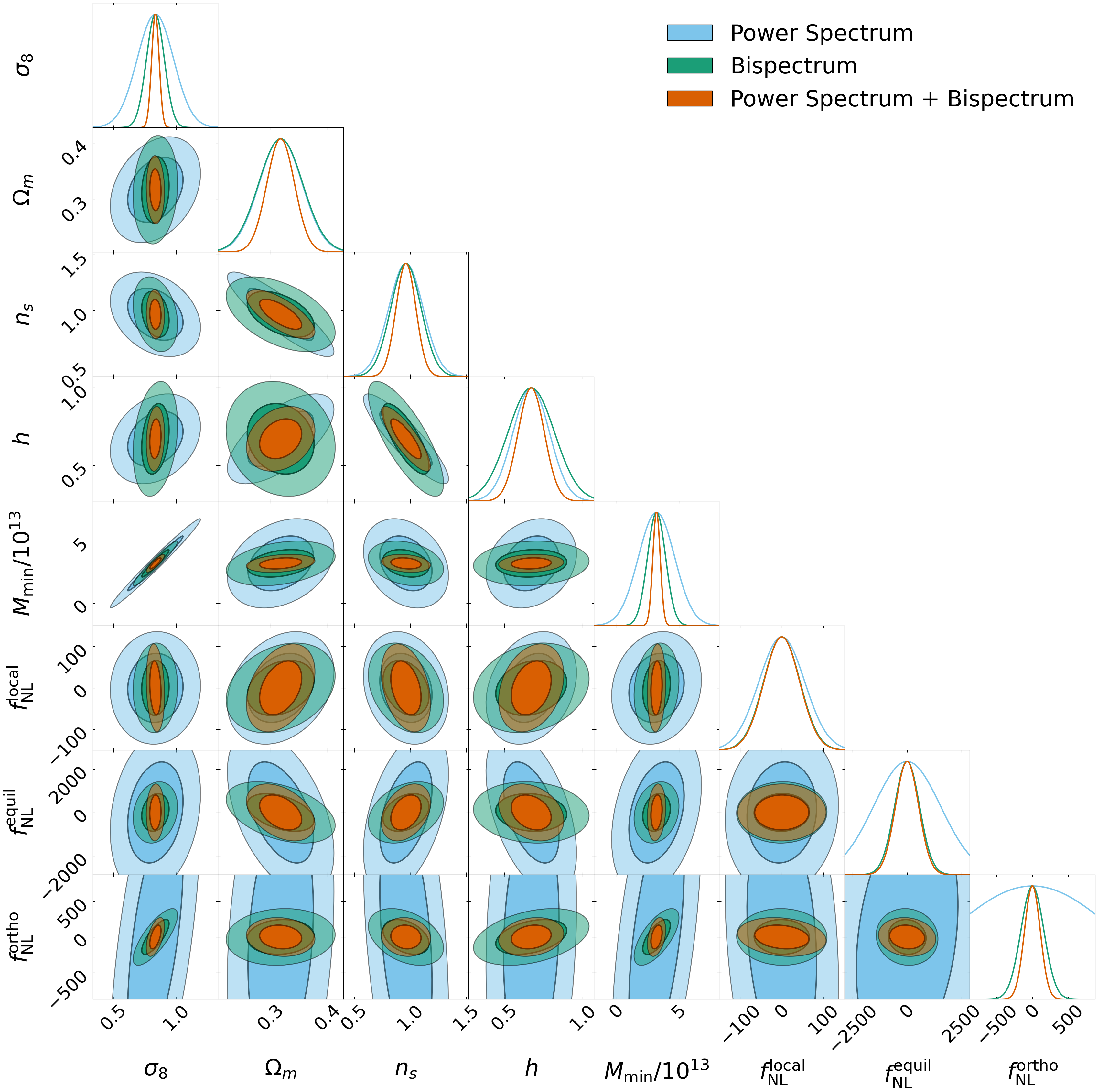}
    \caption{A comparison of the constraining power of the halo power spectrum and bispectrum at $z=1$, for $\kmax0.5~\hMpc$ on cosmological parameters and PNG amplitudes from the power spectrum and the modal bispectrum.}
    \label{fig:fisher_contours}
\end{figure*}

We aim to evaluate the information content on $\Lambda$CDM parameters and PNG amplitudes contained in the power spectrum and bispectrum of the halo field at redshift $z=1$ using a Fisher matrix formalism. This analysis complements the work of \citet{Coulton:2022qbc}, as we focus on a different redshift and make use of a different bispectrum estimator. We explore the dependence on the number of simulations used, the chosen $\kmax$ or the role of the different summary statistics. We show the results in figures~\ref{fig:fisher_compressed}, \ref{fig:fisher_pk_bisp}, \ref{fig:fisher_contours} and table~\ref{tab:constraints}.

As highlighted in \citet{Coulton:2022qbc}, a main difficulty of this simulation-based approach is to accurately compute numerical derivatives of both the power spectrum and bispectrum with respect to the different parameters considered. This is illustrated in figure~\ref{fig:fisher_compressed}, where we show that using smaller subsets of the 500 available pairs of simulations per parameter leads to spurious smaller 1-$\sigma$ uncertainties when analyzing jointly $\{\sigma_8, \Omega_m, h, n_s, M_\mathrm{min}, \fNLloc, \fNLeq, \fNLort \}$ using the traditional Fisher (i.e.\ the dashed lines), indicating a lack of numerical convergence.

Instead of the computationally intensive possibility of producing many more simulations, we use here an alternative method, described briefly in section~\ref{sec:compression} to compute conservative constraints from a lower number of simulations. As expected, the resulting $1$-$\sigma$ error bars decrease when we use more simulations to calculate numerical derivatives, and using the full set they are only between $5\%$ and $25\%$ larger than the unconverged standard Fisher constraints. The results are stable when using $250$ pairs of simulations or more to compute each derivative.

In the simpler situation where we consider only the three following parameters $\{\sigma_8, \Omega_m, \fNLloc \}$ in the analysis, the two methods give very similar results when numerical convergence is reached (using at least 100 pairs of simulations per derivative). This is why in the rest of this work we always use the conservative approach, knowing it is equivalent to the standard Fisher approach in the cases where numerical accuracy can be reached with the available simulations, and otherwise only leads to a reasonable overestimation of order $10\%$ as verified. Note that we manage to keep this overestimation small here due to the use of the modal bispectrum, rather than a standard "binned" approach, because it compresses the original data more efficiently leading to more stable numerical derivatives (we need less than $50$ modes to extract the full information of the bispectrum up to $\kmax=0.5~\hMpc$). This is shown explicitly in appendix~\ref{app:binned}.

In figure~\ref{fig:fisher_pk_bisp} we study the dependence of the constraints on $\kmax$, considering values from $0.1$ to $0.5~\hMpc$. The largest improvement (for both $\Lambda$CDM cosmological and PNG parameters) is obtained between $\kmax=0.1$ and $0.2~\hMpc$, at which point error bars on $\fNLloc$ become saturated (as well as for $h$). However, for the equilateral and orthogonal shapes considering smaller scales yields better constraints (a few percent for each additional increase of $0.1~\hMpc$). For other parameters, the gain can even be larger, justifying the need to probe these nonlinear scales. Note also that all these improvements are computed using the conservative error bars obtained from the compressed summary statistics. Including smaller scales in the analysis typically leads to less converged numerical derivatives, and the less converged these derivatives are the more suboptimal the conservative approach becomes. This means that we may be underestimating slightly the constraining power of the small scales (in any case, this effect should not be large, as for $\kmax=0.5~\hMpc$ this overestimation is $\sim10\%$ as can be seen in figure\ref{fig:fisher_compressed}).  

In figure~\ref{fig:fisher_contours} we show the information content of the halo power spectrum, the halo bispectrum, and their combination. The bispectrum is a much more efficient probe of the equilateral and orthogonal shapes than the power spectrum, while for other parameters they yield constraints of the same order separately. Their combination always helps to reduce degeneracies, although to a lesser extent than for the matter field studied previously in \citet{Coulton:2022qbc, Jung:2022rtn}. 

In table~\ref{tab:constraints}, we present the $1$-$\sigma$ conservative constraints on  $\Lambda$CDM parameters and PNG amplitudes using jointly the power spectrum and bispectrum and including small scales up to $\kmax=0.5~\hMpc$. Unlike the matter field case discussed in \citet{Coulton:2022qbc, Jung:2022rtn}, including PNG shapes in the analysis increases slightly error bars on $\Lambda$CDM cosmological parameters. The different PNG shapes are also less correlated, as analyzing them jointly increases only slightly their own error bars.

\begin{deluxetable*}{l|cccccccc}
 \tablecaption{Joint $1$-$\sigma$ error bars on cosmological parameters and PNG from the power spectrum and the modal bispectrum of the halo field at $z=1$, at $\kmax=0.5~\hMpc$. In the first part, we report the Fisher constraints described in section\ \ref{sec:fisher} and in the second part the corresponding error bars of the quasi-maximum likelihood estimator used in section\ \ref{sec:estimator}. We analyzed 15000 \Quijote\ halo catalogues of $1~(\mathrm{Gpc}/h)^3$ volume at fiducial cosmology, and sets of 500 simulations with one adjusted parameter.
 \label{tab:constraints}}
\tablehead{ & $\sigma_8$ & $\Omega_m$ & $n_s$ & $h$ & $M_\mathrm{min}/10^{13}$ & $\fNLloc$ & $\fNLeq$ & $\fNLort$\\
     Fiducial & 0.834 & 0.3175 & 0.9624 & 0.6711 & 3.2 & 0 & 0 & 0}
\startdata
\hline \hline
Fisher & $\pm 0.024$ & $\pm 0.021$ & $\pm 0.081$ & $\pm 0.078$ & $\pm 0.26$ & & & \\
& $\pm 0.025$ & $\pm 0.022$ & $\pm 0.086$ & $\pm 0.081$ & $\pm 0.27$ & $\pm 42$ & & \\
& $\pm 0.025$ & $\pm 0.023$ & $\pm 0.086$ & $\pm 0.080$ & $\pm 0.27$ & & $\pm 530$ & \\
& $\pm 0.029$ & $\pm 0.021$ & $\pm 0.080$ & $\pm 0.078$ & $\pm 0.29$ & & & $\pm 110$  \\
& $\pm 0.030$ & $\pm 0.025$ & $\pm 0.091$ & $\pm 0.085$ & $\pm 0.29$ & $\pm 43$ & $\pm 530$ & $\pm 110$ \\
\hline \hline
Estimator & $\pm 0.025$ & $\pm 0.022$ & $\pm 0.089$ & $\pm 0.084$ & $\pm 0.27$ & & & \\
& $\pm 0.025$ & $\pm 0.023$ & $\pm 0.098$ & $\pm 0.091$ & $\pm 0.27$ & $\pm 43$ & & \\
& $\pm 0.026$ & $\pm 0.024$ & $\pm 0.092$ & $\pm 0.084$ & $\pm 0.28$ & & $\pm 570$ & \\
& $\pm 0.030$ & $\pm 0.021$ & $\pm 0.085$ & $\pm 0.082$ & $\pm 0.30$ & & & $\pm 110$ \\
& $\pm 0.031$ & $\pm 0.025$ & $\pm 0.094$ & $\pm 0.087$ & $\pm 0.31$ & $\pm 45$ & $\pm 570$ & $\pm 110$ \\
\hline
\enddata
\end{deluxetable*}

\subsection{Parameter estimation}
\label{sec:estimator}
 
As was shown in \citet{Jung:2022rtn} for the matter field, the simple quasi maximum-likelihood estimator (see eq.\ \ref{eq:estimator}) built from the Fisher matrix at some chosen fiducial cosmology is very efficient to measure $\Lambda$CDM cosmological parameters and PNG amplitudes using the power spectrum and bispectrum information. Here we extend this conclusion to the halo field.

The key ingredient of the estimator is the Fisher matrix, which in this work is fully evaluated from a very large set of simulations. As discussed in the previous section, we use a two-step conservative approach for its computation leading to slightly suboptimal results, because numerical convergence is difficult to reach with the standard method. We verify that this leads nonetheless to unbiased and near-to-optimal measurements of parameters, by estimating jointly $\sigma_8$, $\Omega_m$, $n_s$, $h$, $\fNLloc$, $\fNLeq$ and $\fNLort$ in the \Quijote\ simulations using both the power spectrum and the bispectrum.

\begin{figure*}
    \includegraphics[width=0.99\linewidth]{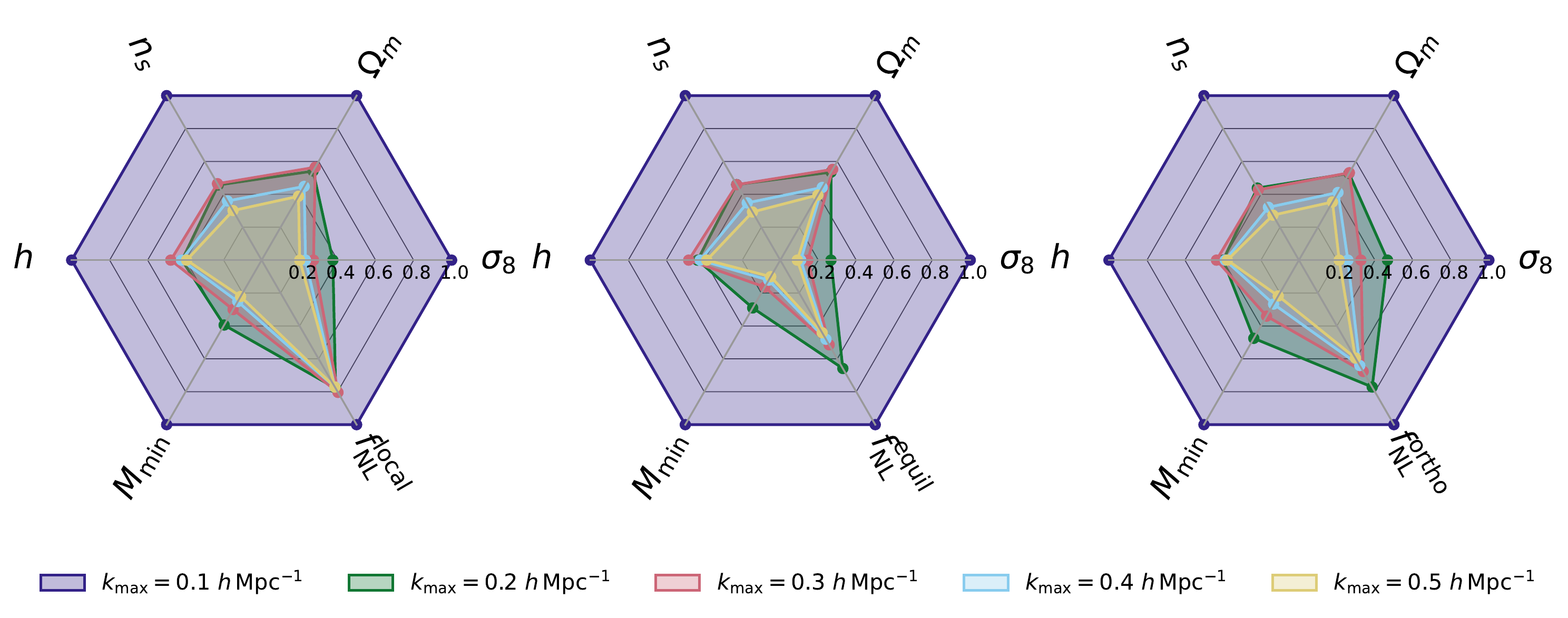}
    \caption{Similar to figure \ref{fig:fisher_pk_bisp}, showing $1$-$\sigma$ error bars of the quasi-maximum likelihood estimator instead of Fisher constraints.}
    \label{fig:errors_pk_bisp}
\end{figure*}

In figure~\ref{fig:errors_pk_bisp}, we study the effect of varying $\kmax$ on the $1$-$\sigma$ error bars of the quasi-maximum likelihood estimator. To compute these error bars, we use a set of $1000$ simulations at fiducial cosmology and analyze it with the estimator calibrated using all other simulations (using $14000$ simulations instead of $15000$ to calculate the covariance has been verified to have no impact on the results). We repeat the procedure for different sets of $1000$ simulations, and compute the standard deviation of the results. As expected, this highlights a very similar behaviour as for the Fisher constraints discussed in the previous section. Concerning PNG, there is no improvement for $\fNLloc$ above $\kmax=0.2~\hMpc$, while for the other two shapes there is no clear saturation yet (although the gain between $\kmax=0.4$ and $0.5~\hMpc$ is only a few percent). For every other parameter considered (except $h$), the decreasing of error bars is significant up to $\kmax=0.5~\hMpc$. In table~\ref{tab:constraints}, we report the corresponding error bars at $\kmax=0.5~\hMpc$, considering cosmological parameters only or jointly with the PNG shapes. For all parameters, the error bars of the quasi-maximum likelihood estimator are close to, or slightly larger than the Fisher constraints reported in the same table (less than $10\%$ difference).

\begin{figure*}
    \includegraphics[width=0.99\linewidth]{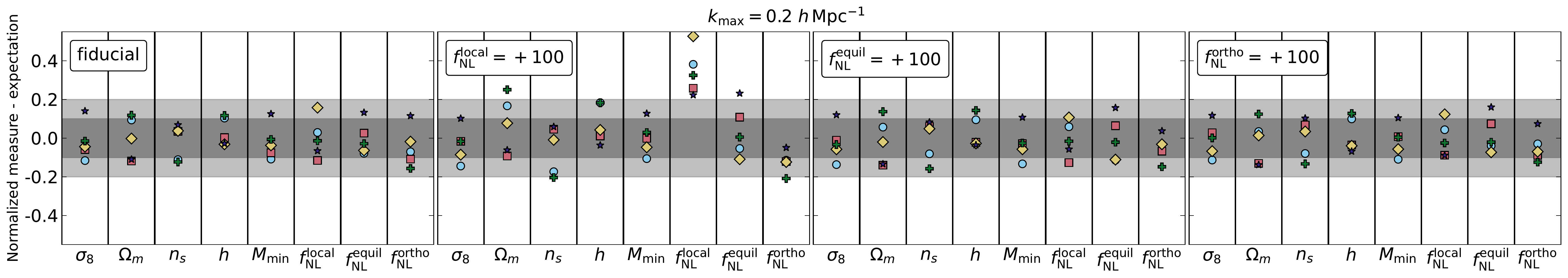}
    \includegraphics[width=0.99\linewidth]{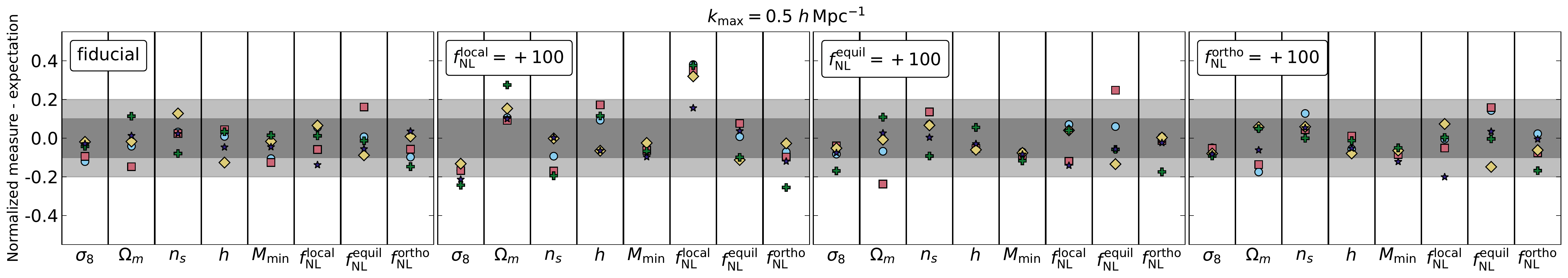}
    \caption{Relative difference of measured cosmological parameters and PNG amplitudes using the quasi maximum-likelihood estimator (eq.\ \ref{eq:estimator}) with their expected value. We use the power spectrum and the bispectrum of the halo field jointly, for $\kmax=0.2~\hMpc$ in the top row and $\kmax=0.5~\hMpc$ below. Each column corresponds to a given parameter (cosmological or PNG). Each panel corresponds to a different input cosmology of the data samples (i.e. one with Gaussian initial conditions and the three types of PNG). For each input cosmology, we analyze five independent datasets of $100$ realizations, each being indicated by its own colour and marker. The dark and light grey bands represent, respectively, the $2$ and $1$-$\sigma$ intervals around the expected deviation ($0$).}
    \label{fig:measures}
\end{figure*}

\begin{figure}
    \includegraphics[width=0.99\linewidth]{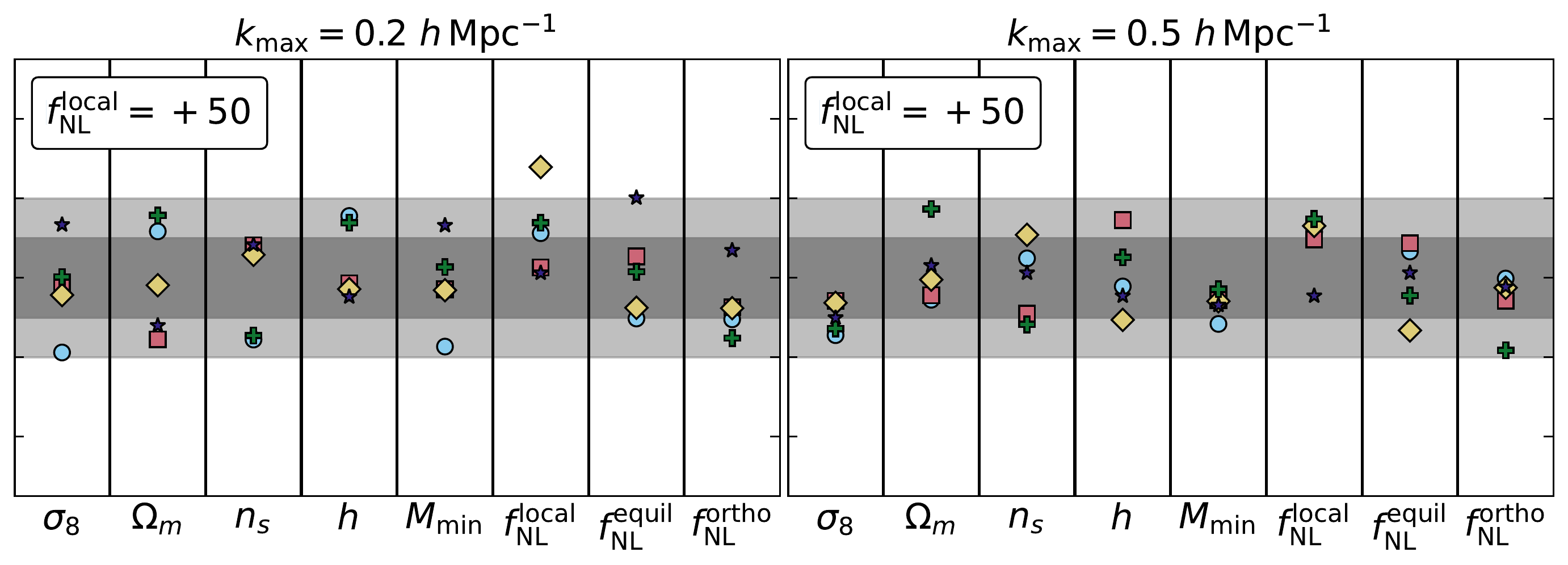}
    \caption{Similar to figure \ref{fig:measures}, for simulations with $\fNLloc=+50$.}
    \label{fig:measures_local}
\end{figure}

In figure~\ref{fig:measures}, we compare the estimated parameters to their input values for different cases, focusing here on changes of PNG amplitudes. We first study the mildly nonlinear regime ($\kmax=0.2~\hMpc$) and then include also nonlinear scales ($\kmax=0.5~\hMpc$). The measured parameters match their expected values for both ranges of scales when studying datasets at fiducial cosmology or having PNG of the equilateral or orthogonal types ($\fNLeq=+100$ or $\fNLort=+100$). 

There are however large statistical deviations on several parameters for the simulations with local NG  (in particular $\fNLloc$, several datasets giving a value more than $5$-$\sigma$ away from the expected one). This difference of behaviour between this specific set and the others can be explained, by the fact that $\fNLloc=100$ is more than $2$-$\sigma$ away from the fiducial value of $\fNL=0$ (based on error bars given in table~\ref{tab:constraints}) while $\fNLort=100$ and $\fNLeq=100$ are respectively smaller and a few times smaller than a $1$-$\sigma$ deviation from $\fNL=0$. The NG simulations of the three shapes correspond to different regimes where a parameter is more or less displaced from the model we use to calibrate the estimator. This is confirmed in figure~\ref{fig:measures_local}, where we check that simulations with $\fNLloc=50$ (thus roughly a $1$-$\sigma$ deviation) give this time the expected results. 

These tests confirm the unbiasedness of the quasi maximum-likelihood estimator, with the caveat that the estimator must be calibrated relatively close to the actual parameter values. This, of course, is due to the fact that the entire method is based on a linear approximation of the likelihood around the fiducial parameters. For the same reason, however, it is clear that the issue can be immediately addressed -- at the computational cost of producing new sets of simulations -- by implementing a standard recursive procedure, in which the estimated parameters at the previous step generate the new fiducial model for the following step, until convergence. Note that this scenario is not bound to occur in practice, since current cosmological parameter constraints from, e.g., CMB datasets such as {\tt Planck} produce already quite narrow priors.
 
\begin{figure}
    \includegraphics[width=0.99\linewidth]{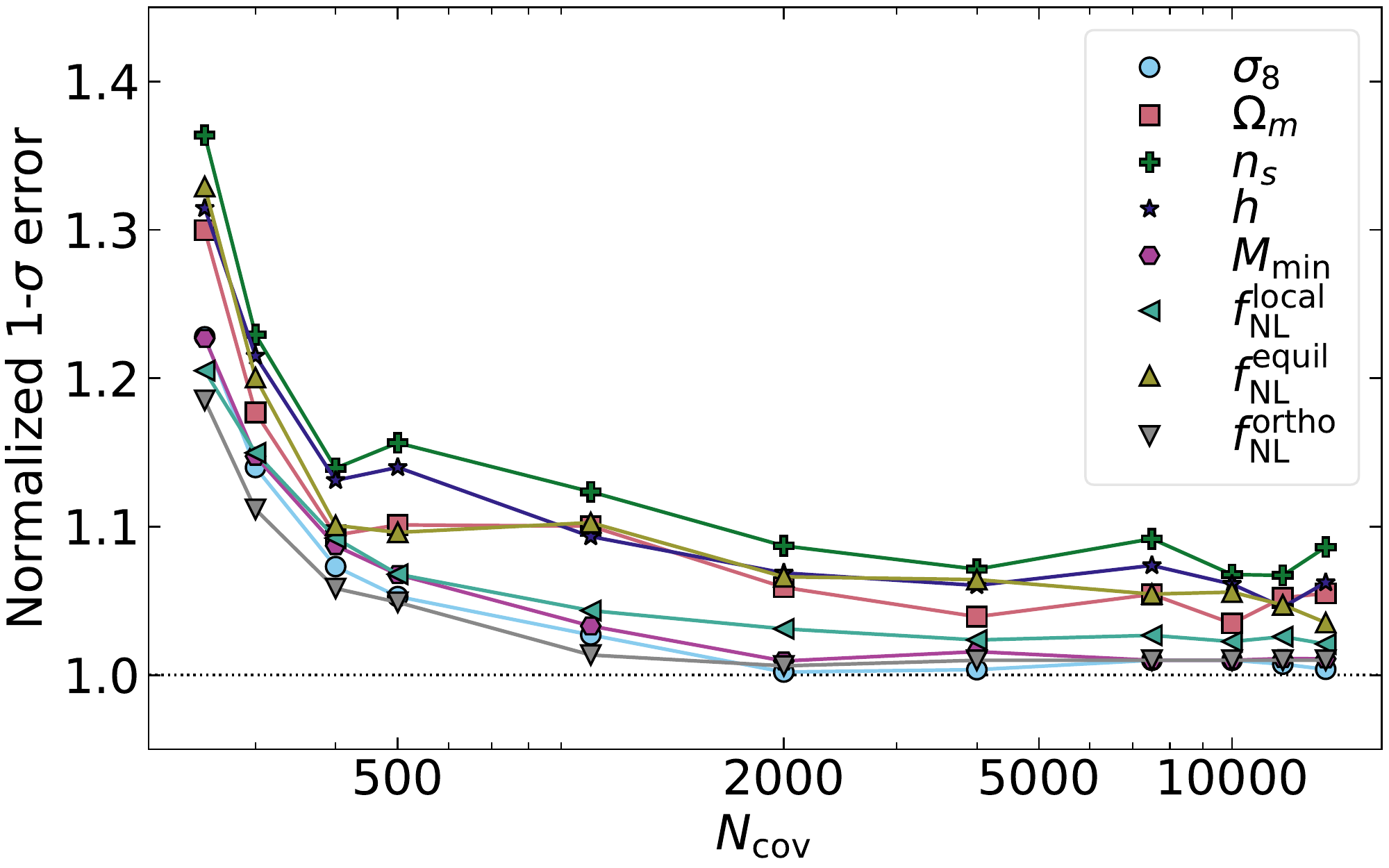}
    \caption{The impact of the number of simulations used to compute the covariance on the error bars of the quasi maximum-likelihood estimator (normalized by the Fisher constraints of table~\ref{tab:constraints}). We use the quasi maximum-likelihood estimator calibrated using all the available \Quijote\ simulations at $z=1$, including the power spectrum and bispectrum information up to $\kmax=0.5~\hMpc$.}
    \label{fig:suboptimality}
\end{figure}

While it was shown in section~\ref{sec:fisher} that using a lower number of simulations to compute derivatives leads to more suboptimal Fisher matrices, it is also important to verify the effect of changes in the number of simulations used to compute the covariance matrix. We explore this in figure \ref{fig:suboptimality}, where we show the increase of error bars due to using fewer simulations. Above 1000 simulations, error bars of the quasi-maximum likelihood estimator are stable (variations at the percent level) and close to the Fisher estimates ($10\%$ difference at most).

\section{Conclusions}
\label{sec:conclusion}

In this paper, we have developed a joint power spectrum and bispectrum quasi-maximum likelihood estimator of cosmological and PNG parameters and applied it to the study of the halo field in the \QuijotePNG\ simulation suite. The data analysis pipeline applies the optimal data compression methodology developed in \citet{Alsing:2017var,Heavens:1999am} to a set of power spectrum and modal bispectrum summary statistics, efficiently extracted from the input mock realizations. In this way, we extended our previous analysis \citep{Jung:2022rtn}, which considered the matter field in the same dataset.

The main arising technical complication was related to the convergence of numerical derivatives that are used to compute the Fisher information and to perform the final compression step. This turns out to be much slower now, with respect to the previous matter field analysis, now leading to potential problems such as spurious "super-optimal'' error bars in the final estimator. Interestingly, though, we have also found that our modal decomposition of the bispectrum makes derivative convergence much faster with respect to the binning approach we implemented in \citet{Coulton:2022rir}. Although still not sufficient for a brute force computation with the available realizations, such faster convergence suggests that more investigation should be done in the future to find the optimal bispectrum decomposition scheme, for the best numerical stability. In the meantime, to circumvent the issue, we have implemented the method first described in \citet{Coulton:2022rir}, which is based on computing the Fisher matrix of MOPED-compressed statistics, extracted from an independent simulation set. This approach leads to stable, robust results, at the price of slight suboptimality in the final estimator.  
Despite such small suboptimality, we have verified that the forecasted errors significantly improve after including non-linear scales up to $\kmax = 0.5~\hMpc$ (see figure \ref{fig:fisher_pk_bisp} as a summary of our main results), in agreement with our findings in \citet{Coulton:2022rir}. Given the significant contribution provided by small scale, shot-noise dominated, bispectrum triangles, further improvements could be in principle achieved in a future galaxy density analysis, by selecting higher-density tracers.
In contrast to other parameters, we have observed a saturation of the $\fNLloc$ error at a scale $k \sim 0.2~\hMpc$; this is again consistent with our previous findings and with other forecasts, such as those in \citet{Karagiannis:2018jdt}, where it was shown that the $\fNLloc$ signal is dominated by the scale-dependent bias signature, on large scales, both in the power spectrum and in squeezed bispectrum configurations.

After investigating the power spectrum and bispectrum information content on non-linear scales, the final step of our analysis consisted in testing our quasi-maximum likelihood estimator on the simulated dataset. We have verified that we can recover unbiased results, deep into the non-linear regime, up to $\kmax = 0.5~\hMpc$ (see figures~\ref{fig:measures} and \ref{fig:measures_local}). Unbiasedness is of course verified only provided the starting fiducial parameter values in the estimator are close enough to the real ones. We studied this in more detail by varying the input value of $\fNLloc$ in the analyzed simulations and verifying that biased results are obtained when the true $\fNLloc$ in the data is $\sim 2 \sigma$ away from the fiducial choice in the estimator. In a realistic observational scenario, this issue can of course always be addressed by implementing a recursive estimation procedure, which however becomes more and more expensive, by requiring new mock realizations and re-calibration of the estimator weights at each step. This suggests to investigate the possibility to reduce the overall computational cost of simulations. We have started a preliminary analysis in this direction, using the CARPool method \citet{Chartier:2020pmu}, which is further discussed in Appendix \ref{app:carpool}. Another possibility is to use machine-learning-augmented simulations, see \citet{Kaushal:2021hqv, Jamieson:2022lqc, Piras:2022dgt} for examples. Making use of these different techniques will play a key role in enabling simulation-based inference with the upcoming generation of galaxy surveys, which will have a much higher tracer density.

The recovered error bars are, as expected, slightly larger than the optimal Fisher bound. This is a direct consequence of the fact that, to secure unbiasedness and robustness of the results, we have calibrated the estimator weights using the stable, yet conservative approximation of the Fisher matrix described above. Also in this case though, the slight suboptimality does not prevent us from obtaining large improvements in precision for the final parameter estimates, when we include non-linear scales in the analysis (see figure~\ref{fig:errors_pk_bisp}).    
By extending our previous analysis to the halo field in redshift space, we have made a significant step forward toward the final development of an efficient, joint power spectrum and bispectrum estimation pipeline, able to extract cosmological and PNG parameters at strongly non-linear scales from actual observations. In a follow-up work we will further extend the current analysis, by looking at the galaxy density field, simulated via a suitable Halo Occupation Distribution (HOD), following \citet{Hahn_Molino}. Marginalization over HOD parameters will also allow us to significantly improve the accuracy of our bias model, which is currently defined by a single parameter which describes the leading order contribution and only to a minor extent captures higher order effects.

Our conclusions are in full agreement with those in our companion work, \citet{Coulton:2022rir}, where we performed an independent analysis at a different redshift \citep[$z=0$ in][vs.\ $z=1$ in this paper]{Coulton:2022rir} and used a standard binning scheme for the bispectrum, rather than the modal approach developed here. Besides increasing the robustness of our conclusions via cross-validation of independent data analysis pipelines, the two works complement each other in several ways and together cover a significant range of crucial issues: \citet{Coulton:2022rir} focused on addressing numerical stability issues, on assessing the information content of our observables at different scales and on evaluating in detail all possible contributions to the error budget (such as, e.g., shot noise and super-sample covariance effects), whereas the present study, while cross-checking the previous Fisher matrix results, is more centered on optimal data compression and on the development and testing of related statistical estimators.

\section*{Acknowledgements}

\noindent GJ, ML and MB were supported by the project "Combining Cosmic Microwave Background and Large Scale Structure data: an Integrated Approach for Addressing Fundamental Questions in Cosmology", funded by the MIUR Progetti di Ricerca di Rilevante Interesse Nazionale (PRIN) Bando 2017 - grant 2017YJYZAH. 

\noindent DK is supported by the South African Radio Astronomy Observatory (SARAO)
and the National Research Foundation (Grant No. 75415).

\noindent GJ, and ML also acknowledge support from the INDARK INFN Initiative (\url{https://web.infn.it/CSN4/IS/Linea5/InDark}), which provided access to CINECA supercomputing facilities (\url{https://www.cineca.it}).

\noindent MB acknowledges the use of computational resources from the parallel computing cluster of the Open Physics Hub (\url{https://site.unibo.it/openphysicshub/en}) at the Physics and Astronomy Department in Bologna.

\noindent LV acknowledges  ERC (BePreSySe, grant agree- ment 725327),  PGC2018-098866- B-I00 MCIN/AEI/10.13039/501100011033 y FEDER “Una manera de hacer Europa”, and the “Center of Excellence Maria de Maeztu 2020-2023” award to the ICCUB (CEX2019-000918-M funded by MCIN/AEI/10.13039/501100011033).

\noindent B.D.W. acknowledges support by the ANR BIG4 project, grant ANR-16-CE23-0002 of the French Agence Nationale de la Recherche; and the Labex ILP (reference ANR-10-LABX-63) part of the Idex SUPER, and received financial state aid managed by the Agence Nationale de la Recherche, as part of the programme Investissements d'avenir under the reference ANR-11-IDEX-0004-02.
The Flatiron Institute is supported by the Simons Foundation.


\appendix
\section{Shot noise modal modes}
\label{app:shot-noise}

The shot-noise contribution to the matter bispectrum at tree-level is given by
\begin{equation}
    \label{eq:bispectrum-shot-noise}
    B^\mathrm{SN}(k_1, k_2, k_3) = \frac{1}{\bar{n}} \left[P_L(k_1) + P_L(k_2) + P_L(k_3)\right] + \frac{1}{\bar{n}^2}, 
\end{equation}
where $\bar{n}$ is the halo number density and $P_L(k)$ is the linear matter power spectrum. As introduced in \citet{Byun:2020rgl}, this can be fully described in the modal way by using the two triplets $(0,0,1)$ and $(0,0,0)$ combining the following one-dimensional basis functions
\begin{equation}
        q_0 =\sqrt{\frac{k}{P(k)}} \frac{5}{14}, \qquad q_1 =\sqrt{\frac{k}{P(k)}} P_L(k).
\label{eq:modal-shot-noise}
\end{equation}

\section{Comparison with the standard "binned" bispectrum estimator}
\label{app:binned}

A key ingredient to compute the Fisher matrix (eq.\ \ref{eq:fisher}), which is used both for constraint forecasts and to build estimators, is to have accurate derivatives of the summary statistics with respect to the different parameters considered. As discussed in section~\ref{sec:fisher}, even the large sets of $500$ paired simulations for each parameter of the \Quijote\ and \QuijotePNG\ collections are not sufficient to reach the necessary numerical convergence for the power spectrum and bispectrum derivatives. This typically leads to an underestimation of $1$-$\sigma$ error bars. 

On the other hand, a two-step computation, consisting first on compressing optimally the data, and then computing the Fisher matrix from this compressed data (using different datasets for the two steps), yields slightly overestimated error bars. Combining the power spectrum and the bispectrum information of the halo field at $z=1$, we have verified that even when we include nonlinear scales up to $\kmax=0.5~\hMpc$, the difference between the lower and upper bounds of constraints is at most of order $20\%$ on the different parameters, a very reasonable difference. In this appendix, we show that the modal estimator, which by construction compresses the bispectrum information in the data, is a necessary ingredient for the efficiency of the method. 

\begin{figure*}
    \includegraphics[width=0.99\linewidth]{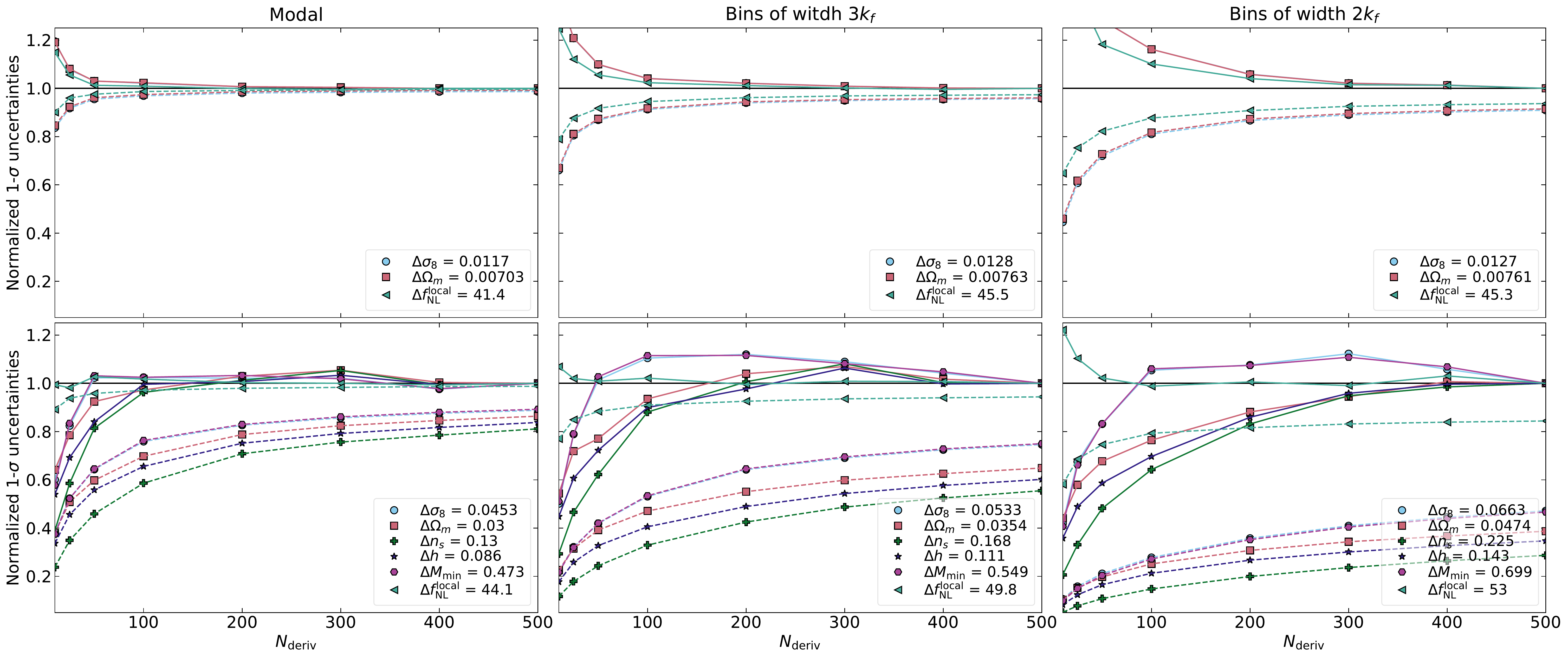}
    \caption{A comparison of $1$-$\sigma$ uncertainties obtained with different bispectrum estimators. The left column is obtained with the modal bispectrum estimator used throughout this paper, while the two others use a standard "binned" approach for different widths of bin ($3k_f$ in the middle panels and $2k_f$ in the right panels). Otherwise, this figure is similar to \ref{fig:fisher_compressed}, for $\kmax=0.2~\hMpc$.}
    \label{fig:binned_modal}
\end{figure*}

In figure~\ref{fig:binned_modal}, we compare the convergence of standard and conservative Fisher $1$-$\sigma$ uncertainties obtained with the modal bispectrum (as in the rest of this paper) and a standard "binned" bispectrum estimator.\footnote{As was pointed out in \citet{Byun:2020rgl}, the standard bispectrum estimator in Fourier space can be recovered in the modal fashion by using a simple basis with modes of the form $q_n(k) = \sqrt{k P(k)}\theta_i(k)$ where $\theta_i(k)=1$ if $k\in\Delta_i$ ,after dividing the $k$-range into bins $\Delta_i$, and $0$ otherwise. This is this implementation we use for this analysis.} A main result is that the constraints obtained with the modal estimator are the most stringent, with a difference of order $10\%$ with the standard estimator with bins of width $3k_f$, and even more with smaller bins of width $2k_f$. Indeed the estimator using the smallest bins gives here the largest error bars, despite the fact that in principle it should keep more information, due to the greater difficulty of computing sufficiently accurate numerical derivatives. This lack of convergence is also very clear when we compare the lower and upper bounds on error bars for all three methods. Using the full sets of simulations, the lower bounds are $10$-$20\%$ smaller than the upper limits for the modal estimator, $30\%$ for bins of width $3k_f$, and as much as two times smaller for bins of width $2k_f$. The modal estimator gives more stringent constraints, which are proven to be closer to the actual Fisher uncertainties, and should converge totally with a smaller number of simulations, as shown in the first row in the simple situation, where the modal estimator has fully converged and the other two have not.

\begin{figure*}
    \includegraphics[width=0.99\linewidth]{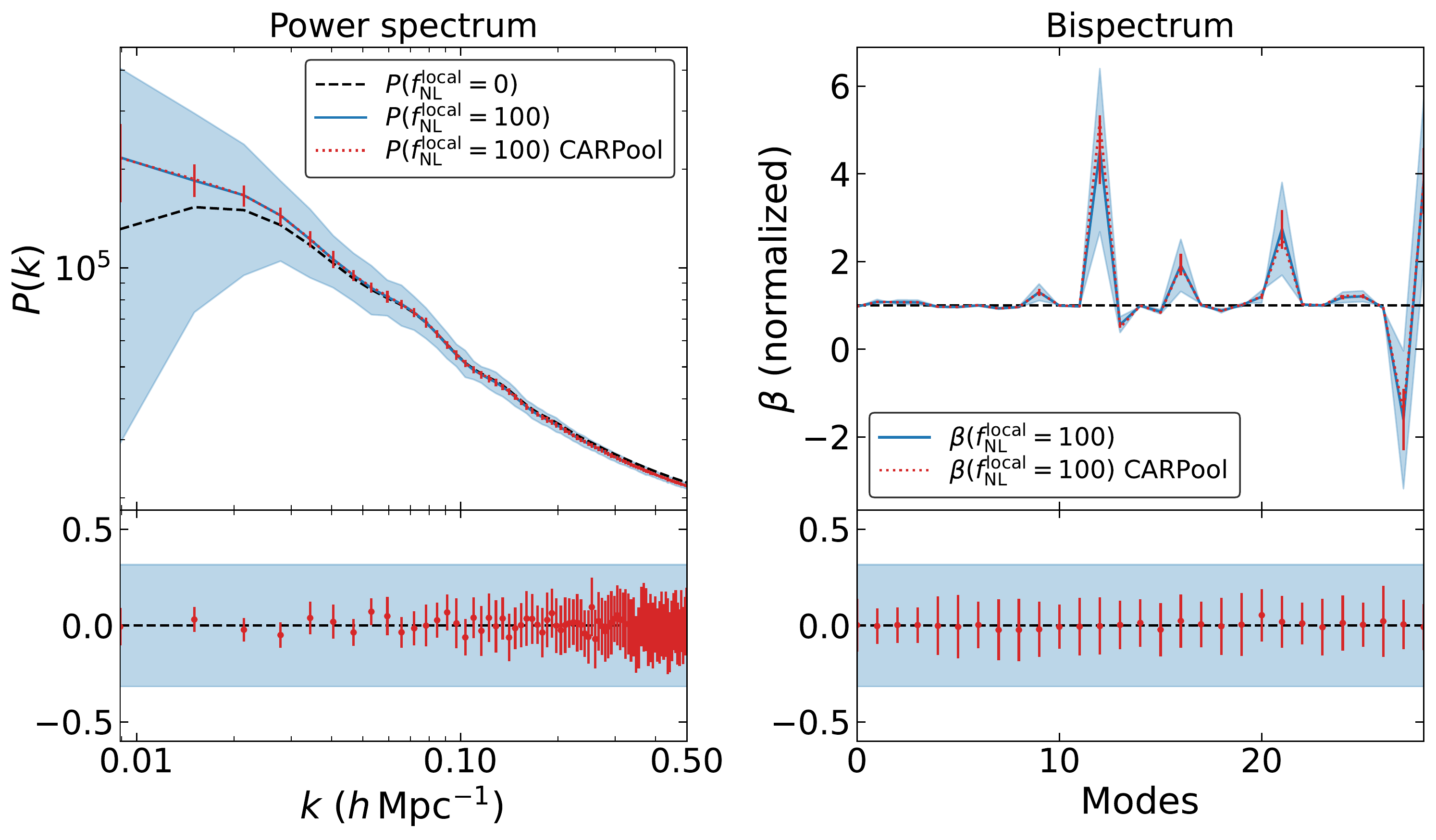}
    \caption{The CARPool method applied to the power spectrum (left column) and modal bispectrum (right column) of the halo field, at $z=1$. On the top row, the black dashed lines correspond to the averages from the $15000$ \Quijote\ simulations at fiducial cosmology  with $\fNLloc=0$ (note that in the bispectrum case, all modal coefficients are normalized by dividing by the modes from these $15000$ simulations at fiducial cosmology). The blue lines correspond to the average from $500$ simulations with $\fNLloc=+100$. The red dotted lines have been computed using the CARPool method (see eq.\ \ref{eq:carpool}), using $10$ simulations at $\fNLloc=+100$ as the high-fidelity simulations and the 15000 simulations at fiducial cosmology as surrogates. The blue areas and red vertical lines show the respective error bars from the two cases (they correspond to standard errors for sets of $10$ simulations, and have been multiplied by a factor $10$ for visibility in the power spectrum case). In the bottom row, we show the difference between the CARPool estimates and averages from $500$ simulations, normalized by the standard deviation. The blue areas correspond to the standard error for $10$ simulations, and error bars. Error bars on the CARPool estimates, shown in red, are computed by applying the CARPool method to many different sets of $10$ simulations at $\fNLloc=+100$.}
    \label{fig:carpool_statistics}
\end{figure*}

\section{Application of CARPool}
\label{app:carpool}

As we have verified in this paper, the quasi-maximum likelihood estimator is a powerful method to infer cosmological parameters and PNG amplitudes from halo catalogues using information beyond the mildly non-linear regime, which however, as other simulation-based methods, can require a large number of costly forward simulations. Therefore, a key component of future applications will be to include the variance reduction CARPool technique, developed in \citet{Chartier:2020pmu, Chartier:2021frd, Chartier:2022kjz}, into the full analysis pipeline.

The basic idea behind CARPool is to use a relatively small number of high fidelity simulations combined with a large number of less accurate simulations, or surrogates, to measure some chosen summary statistics with much smaller error bars. In \citet{Chartier:2020pmu}, these surrogates were computed using much faster, but less precise, N-body solvers like COLA \citep{Tassev:2013pn}. This could for example be applied to the case of numerical derivatives, for which reaching numerical convergence typically requires thousands of costly simulations. We leave this application for future work, and instead focus here on the use of CARPool to speed-up the iteration process of quasi-maximum likelihood estimation.

To obtain unbiased estimates of cosmological parameters or $\fNL$'s, it is important that the fiducial cosmology where we evaluate the covariance and numerical derivatives is not too far from the actual parameter values. For example, in section~\ref{sec:analyses}, we have seen that with a fiducial cosmology at $\fNLloc=0$, the estimator is unbiased in measuring $\fNLloc$ in simulations with an input of $\fNLloc=50$, but not for $\fNLloc=100$. Note that even if the measured bias were large when averaging from hundreds of simulations, it would still be smaller than the $1$-$\sigma$ error bar, making it a good first estimate. Then, working by iteration and choosing a new fiducial cosmology at these roughly measured parameters should yield unbiased results. To avoid producing a completely new large set of simulations at the new fiducial cosmology, we can consider the original simulations as the surrogates of the CARPool method \citep[the idea of combining simulations at different cosmology was also explored in ][]{Ding:2022ydj}.

\begin{figure*}
    \includegraphics[width=0.99\linewidth]{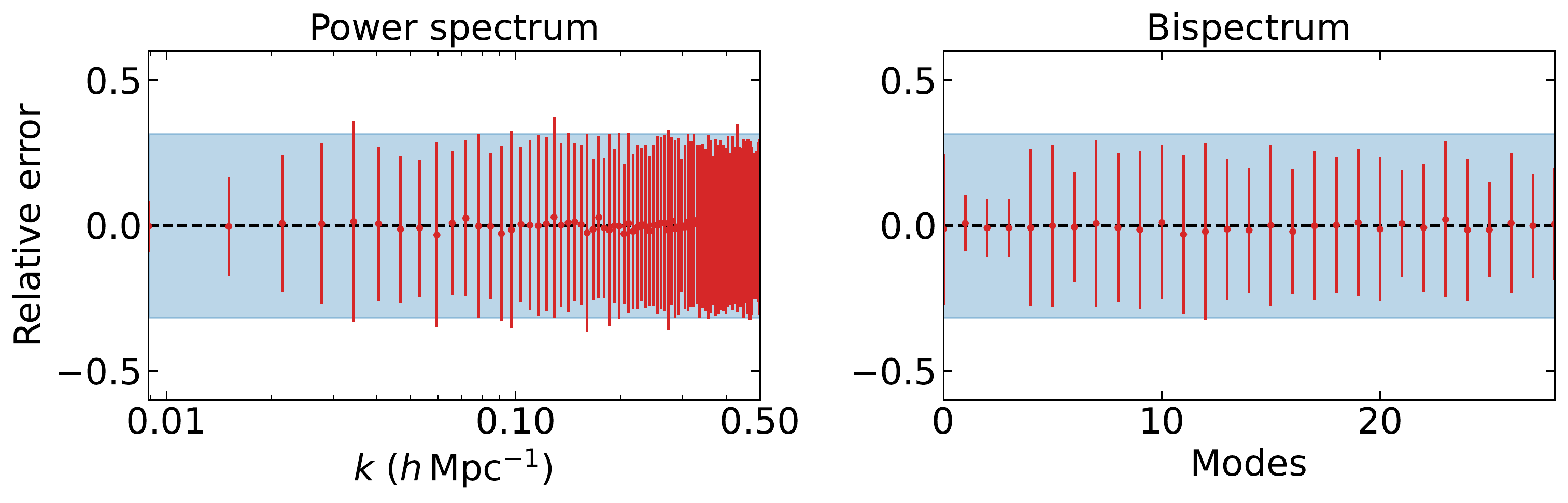}
    \caption{Similar to the bottom line of figure~\ref{fig:carpool_statistics}, where the CARPool method is applied to compute the derivatives $\partial P(k)/\partial\fNLloc$ and  $\partial\beta/\partial\fNLloc$ at the point $\fNLloc=+50$ , using the previously computed derivatives at $\fNLloc=0$ as surrogates. }
    \label{fig:carpool_derivatives}
\end{figure*}

The main ingredients of the CARPool method are:
\begin{itemize}
    \item A set of $N$ paired high-fidelity simulations and surrogates, sharing the same random seeds to produce their initial conditions, from which we measure some chosen summary statistic denoted  $\mathbf{y}$ or $\mathbf{c}$ (simulation or surrogate respectively) and the corresponding sample covariance given by
    \begin{equation}
        \label{eq:carpool-covariance}
        \mathbf{\hat{\Sigma}_{yc}}=\frac{1}{N-1}\sum\limits_{i=1}^{N}(\mathbf{y}_i-\mathbf{\bar{y}})(\mathbf{c}_i-\mathbf{\bar{c}})^\mathbf{T}, \qquad \mathbf{\hat{\Sigma}_{cc}}=\frac{1}{N-1}\sum\limits_{i=1}^{N}(\mathbf{c}_i-\mathbf{\bar{c}})(\mathbf{c}_i-\mathbf{\bar{c}})^\mathbf{T}, \qquad \mathbf{\bar{y}}=\frac{1}{N}\sum\limits_{i=1}^{N}\mathbf{y}_i.
    \end{equation}
    \item A separate set of $M$ surrogates, to compute the mean of $c$ with the standard expression
    \begin{equation}
        \label{eq:carpool-mean}
        \bm{\mu_c}=\frac{1}{M}\sum\limits_{i=1}^{M}\mathbf{c}_i.
    \end{equation}
\end{itemize}
Then, the key quantity to compute is
\begin{equation}
    \label{eq:carpool}
    \mathbf{x} = \mathbf{y} - \bm{\hat{\beta}}(\mathbf{c}-\bm{\mu_c}),
\end{equation}
which by construction has the same ensemble average as $\mathbf{y}$ (i.e.\ $\mathbf{\bar{x}}=\mathbf{\bar{y}}$). The variance of $x$ is minimized when the control matrix $\mathbf{\hat{\beta}}$ is given by 
\begin{equation}
    \bm{\hat{\beta}}=\mathbf{\Sigma_{yc}}\mathbf{\Sigma_{cc}^{-1}}.
\end{equation}
In \citet{Chartier:2020pmu}, it was shown that a very efficient choice, using only the diagonal elements of $\mathbf{\Sigma_{yc}}$ and $\mathbf{\Sigma_{cc}}$, is the following diagonal control matrix:
\begin{equation}
    \label{eq:control}
    \bm{\beta^\mathrm{diag}} =  \mathrm{diag}\left(\frac{\mathrm{cov}(y_1, c_1)}{\sigma(c_1)^2},\frac{\mathrm{cov}(y_2, c_2)}{\sigma(c_2)^2},...,\frac{\mathrm{cov}(y_n, c_n)}{\sigma(c_n)^2}\right),
\end{equation}
where $n$ is the size of the vectors $\mathbf{y}$ and $\mathbf{c}$.

In figure~\ref{fig:carpool_statistics}, we show the results obtained with the CARPool technique applied to the \QuijotePNG\ set of halo catalogues. We use the large set of $15000$ \Quijote\ simulations at fiducial cosmology and with no PNG in their initial conditions as the surrogates, and a small set of $10$ non-Gaussian simulations ($\fNLloc=+100$) with the same $\Lambda$CDM cosmological parameters as the high-fidelity simulations, the goal being to predict the power spectrum and bispectrum more accurately outside of fiducial cosmology. We compare the CARPool results to the 500 simulations with $\fNLloc=+100$ at our disposal and verify that they are indeed unbiased, as expected. We repeat the procedure to many $10$ simulation subsets among the $500$ to check that the result is not spurious, and to derive error bars on the CARPool averages. For all power spectrum and bispectrum modes, the error bars are significantly smaller than the standard errors on the average from $10$ simulations alone. The effect is the strongest on linear scales (small $k$ for the power spectrum, and the first few bispectrum modes which describes the tree-level matter bispectrum), but is also present in the non-linear regime.

In figure~\ref{fig:carpool_derivatives}, we follow a similar procedure to study the derivatives of the power spectrum and bispectrum with respect to $\fNLloc$. We use the derivatives evaluated at $\fNLloc=0$ by finite difference applied to the $500$ $\fNLloc=\pm100$ simulations as surrogates, to compute the derivatives at $\fNLloc=50$ using only a few simulations with $\fNLloc=0$ or $100$. For the power spectrum the improvement is small, or even negligible in some cases, outside of the largest scales. For the bispectrum, there is a significant improvement of the first few modes describing the tree-level matter bispectrum. All error bars are reduced by the CARPool method, although the improvement is very small for some modal coefficients. One issue here is the small number of surrogates compared to the previous application (only $500$ instead of $15000$), adding to the fact that we know that the surrogate derivatives are not even fully converged numerically (as discussed thoroughly in section~\ref{sec:analyses}).

These examples illustrate briefly the possibilities of the CARPool technique. We leave its full implementation in the pipeline for future works, where we will also include the powerful "CARPool Bayes" introduced in~\citet{Chartier:2022kjz} for the fast and accurate estimation of the covariance matrix and its inverse.

\bibliographystyle{aasjournal}
\bibliography{biblio}

\end{document}